\documentclass[preprint,12pt]{elsarticle}
\usepackage{amsmath}
\usepackage{amsfonts}
\usepackage{amssymb}
\usepackage[utf8]{inputenc}
\usepackage[T2A]{fontenc}
\usepackage{bm}
\usepackage{graphicx}

\begin{document}

	\title{Soliton radiation and role of the energy dissipation \\
		in soliton dynamics in an oscillating magnetic field} 
	
	\author{Larissa Brizhik}
	\address{Bogolyubov Institute for Theoretical Physics of the National Academy of Sciences of Ukraine \\
		Metrologichna Str., 14-b,  Kyiv, 03143, Ukraine.	\\[0pt]
		\vspace*{0.25cm}
		brizhik@bitp.kyiv.ua }
	
	\begin{abstract}
		{Dynamics of the Davydov’s soliton in an external oscillating in time magnetic field is studied analytically. It is shown that in a field perpendicular to the molecular chain axis, soliton wavefunction is a product of the electron plane wave in the plane perpendicular to the molecular chain, and longitudinal component of the wavefunction which satisfies the modified Nonlinear Schr\"odinger equation with an extra term determined by the field. It is shown that soliton  width and amplitude are constant, while its velocity and phase are oscillating functions of time with the frequency of the main harmonic equal to the magnetic field frequency. It is shown that soliton dynamics has two different regimes at low and high frequencies of the magnetic field as comparing with the characteristic soliton frequency.  Due to time-depending velocity and nonzero acceleration, soliton radiates linear waves in both directions from its center of mass. In the presence of the energy dissipation, soliton velocity is bound from above due to the balance of the energy gain from the magnetic field, and its loss because of the  dissipation and radiation of linear sound waves. This balance occurs at the resonant frequency of the magnetic field. It is concluded that such significant impact of time-depending magnetic fields on charge transport, provided by solitons, can affect functioning of the devices based on low-dimensional molecular systems. These results suggest the physical mechanism of therapeutic effects of oscillating magnetic fields. 
		}
		
	\end{abstract}
	
	\maketitle

\section{Introduction}

In 1973,  Olexandr Davydov and Mykola Kyslukha suggested the mechanism of the energy transfer on macroscopic distances in biological macromolecules, based on the account of the electron-lattice interaction \cite{DavydovKislukha1}. They have shown that due to this interaction the bound state of a molecular excitation, namely, Amid-I vibration in the $\alpha-$helical protein, is self-trapped in the potential well of the local deformation of the molecular chain, and together they form a nonlinear bound state, now known as the Davydov's soliton \cite{Scott,Davydov}.  Let us recall, first, that soliton is a solution of the nonlinear differential equation(s) in the form of a solitary wave localized in space, which can move without energy dissipation and its speed and amplitude remain unchanged under the interaction with a similar, but not necessarily identical,  wave. In particular, the Davydov's soliton is formed due to the polaron effect and in continuum representation in the adiabatic approximation is described by the nonlinear Schr\"odinger equation. The variational study \cite{ground} has shown that solitons correspond to the ground electron states in molecular  systems  with intermediate values of the electron-lattice interaction,  while in systems with strong electron-lattice coupling the ground electron state is described by small polaron, and in systems with  very weak electron-lattice coupling  electrons are in an almost free state. Therefore, Davydov's solitons correspond to crossover  between strong and weak electron-phonon coupling regimes and can be compared with large polarons with the principal difference, that in solitons this coupling is taken into account without using perturbation theory or linearization. The soliton model found numerous applications in physics and biophysics and was generalized to describe also charge transport and to take into account many other factors, such as more realistic structure of molecular chains, energy dissipation, impact of temperature,  etc. (see \cite{Scott,Davydov}). 

There exists a large class of low-dimensional molecular systems many of which due to their unique properties are important as functioning materials for micro- and nano-electronics. These systems  include biological macromolecules, organic and inorganic systems, such as polydiacetylene, conducting polymers, some superconducting compounds and many other, in which electron-lattice interaction is sufficiently strong and leads to self-trapping of quasiparticles (molecular excitations, extra electrons or holes) in soliton states, providing, thus, the efficient mechanism of charge and energy transport on macroscopic distances (see, e.g.,  \cite{Brizhik-DAA} and references therein). In real conditions these systems are often exposed to external fields, and, in particular, to magnetic fields (MFs), for instance, from technical appliances or generated by the inner parts of the devices. This subject  is important also from the point of view of biological systems, since such  fields can affect the redox processes in living organisms.  Moreover, weak oscillating MFs are used in therapies for treatment of various diseases, although little is known about  physical mechanism of these therapies  \cite{BrizhikArXiv,BrizhikFermiZavan,BrizhikFermi}. As it is well known, practically all diseases are accompanied by oxydative stress, related to the violations of the redox processes in mitochondria. Therefore, we can expect, that oscillating MFs, changing the dynamics of solitons in macromolecules in the electron transport chain of the Krebbs cycle \cite{BrHoMu}, can affect the redox processes. As it will be shown below, this impact has a resonant character, and, thus, this can constitute the mechanism of the resonant magnetic field impact on biological systems in addition to other linear mechanisms (see, e.g., \cite{McLeod}). 

Soliton dynamics in constant magnetic field has been studied earlier \cite{BrizhikMF1,BrizhikMF2}. In \cite{MF-Arxiv,MF-SChFr} impact of oscillating in time magnetic field on soliton dynamics was studied in the zero adiabatic approximation. In this paper this problem is studied with account of the energy dissipation and soliton radiation will be calculated. It will be shown that such fields have a resonant impact on soliton dynamics. Soliton parameters attain time dependence. In particular, soliton width and acceleration are oscillating functions of time with the frequency of the main harmonic determined by the frequency of the field. Due to such dynamics, solitons emit linear waves in both directions from its center of mass. This process is the most intensive at the resonant frequency of the magnetic field determined by the characteristic soliton frequency. It will be shown also that energy dissipation plays stabilizing role in soliton dynamics in magnetic fields. The paper is organized as follows. In Section \ref{Dav-sol} a brief description of  Davydov's soliton will be given and soliton dynamics in an  oscillating MF will be discussed  (Section \ref{MF}). In Section \ref{deform} it will be shown that in the self-consistent description of electron and lattice deformation solitons have two different dynamic regimes at low and high frequency MFs and in Section \ref{dissip} significant role of  energy dissipation in soliton dynamics will be  demonstrated. Finally, in Conclusions, we will summarize the obtained results and their potential role for functioning of nanoelectronic devices, based on novel functional low-dimensional compounds. We will also discuss impact of change of soliton dynamics in MF in biological systems.

\section{Davydov's soliton}\label{Dav-sol}

Here we briefly describe the model (detailed description of the Davydov's soliton in a molecular chain with the electron-lattice interaction can be found in \cite{Davydov}). Consider a one-dimensional molecular chain with an extra electron and take into account electron-lattice interaction. Such a system is described  by the Fr\"ohlich Hamiltonian
\begin{equation}
	H_F={{H}_{el}}+{{H}_{latt}}+{{H}_{int}},
	\label{HF}
\end{equation}
where the terms ${H}_{el}$, ${H}_{latt}$ and ${H}_{int}$ are the Hamiltonians of the electron, lattice vibrations (phonons) and electron-lattice interaction. From this Hamiltonian one can derive a
nonlinear system of equations for the electron wavefunction ${\psi (x, t)}$ normalized to unity, 
 and lattice deformation ${\rho (x, t)}$, which for our further purpose  we write down in the continuum representation \cite{Davydov}
\begin{equation}
	\label{eqpsi1}
	i\hbar \frac{\partial }{\partial t}\psi (x, t)=\left[ \frac{p_x^2}{2m_x}-\sigma \rho(x,t)\right]\psi (x,t),
\end{equation}
\begin{equation}
	\label{eqrho1}
	\left( \frac{{{\partial }^{2}}}{\partial {{t}^{2}}}-{{V}_{0}}^{2}\frac{{{\partial }^{2}}}{\partial {{x}^{2}}} \right)\rho (x, t)+\frac{\sigma {{a}^{2}}}{M}\frac{{{\partial }^{2}}}{\partial {{x}^{2}}}{{\left| \psi (x, t) \right|}^{2}}=0.
\end{equation}
Here  $x$ is the coordinate along the chain,  $p_x=i\hbar \partial  /\partial x $ is the electron momentum,  ${\sigma}$ is the electron-lattice coupling constant,  ${a}$ is the lattice constant, $m_x=\hbar ^2 /2Ja^2$ is the effective electron mass, which is determined by the exchange interaction $J$,  ${M}$ is the mass of a molecule,  ${{V}_{0}}$ is the sound velocity in the chain. 

In the adiabatic approximation the solution of Eq. (\ref{eqrho1}) is 
\begin{equation}
	\label{rho}
	\rho(x,t)=\frac{ \sigma }{w(1-s^2)}|\psi (x,t)|^2,  
\end{equation}
and Eq. (\ref{eqpsi1}) is transformed to the nonlinear Schr\"odinger equation (NLS)
\begin{equation}
	\label{NLS}
	\left( i\hbar \frac{\partial }{\partial t}+Ja^2\frac{{{\partial }^{2}}}{\partial {{x}^{2}}}+2Jg |\psi (x,t)|^2 \right)\psi (x,t)=0,
\end{equation}
with the well-known soliton solution
\begin{equation}
	\label{psi-sol}
	\psi (x,t)=\psi _s(x,t)\equiv \frac{\sqrt{g} }{2} \operatorname{sech} \left[ g(x-x_0-Vt)/a \right] e^{  i({{k}_{x}}x-\omega_s t) }
\end{equation}
in which ${g}$ is the dimensionless nonlinearity constant
\begin{equation}
	\label{g}
	g=\frac{ \sigma ^2}{2Jw(1-s^2)}, \qquad s^2=\frac{V^2}{V_0^2},
\end{equation}
$V$ is soliton velocity, $w=MV_0^2/a^2$ is the chain elasticity, $k_x=m_xV/\hbar$ is the $x$-component  of the electron wave-vector, $\omega_s $ is the frequency determined by the eigen-energy of the soliton, $x_0$ is the soliton center of mass position at the initial time moment $t=0$ which for simplicity can be set  equal to zero.

\section{Soliton dynamics in a magnetic field}\label{MF}

To study soliton dynamics in the external MF, we have to generalize the model to a three-dimensional case. We can represent the wavefunction as the product 
\begin{equation}
	\label{psi3D}
	\Psi (\vec{r}, t)=\psi (x,t)\psi_{tr} (y,z,t)
\end{equation}
and take into account that in the presence of the MF, the  electron momentum $ {\vec p} ({\vec r})=i\hbar\partial / \partial {\vec r} $ has to be modified as follows
\begin{equation}
	\label{mom}
	{\vec p}({\vec r}) \rightarrow	{\vec P}({\vec r})\equiv  {\vec p}({\vec r}) - \frac{e}{c}{\vec A} 
\end{equation}
where  ${\vec A}$ is the MF vector-potential  ${\vec B}=\rm{rot}\vec{{A}}$, $e$ is the  electron charge and $c$ is the speed of light. 

Therefore, the electron wavefunction in the external MF is determined by the equation
\begin{equation}
	\label{psi}
	i\hbar \frac{\partial }{\partial t}\Psi (\vec{r}, t)=H\Psi (\vec{r}, t),
\end{equation}
with the Hamiltonian   
\begin{equation}
	\label{H}
	H=\sum\limits_{\nu =x,y,z}{\left[ {{\left( {{p}_{\nu }}-\frac{e}{c}{{A}_{\nu }} \right)}^{2}}\frac{1}{2{{m}_{\nu }}}-\sigma \rho (x, t) \right]}
\end{equation}
where ${{m}_{\nu}}$ are the components of the effective electron mass in the conduction band of the system, $\nu = x,y,z$.

For our study it is useful to represent the harmonic MF ${\vec B}(t)=\vec {B}_0\cos (\omega t)$ where $B_0$ is the intensity of the field and $\omega $ is its frequency,  via its longitudinal and perpendicular to the chain orientation components. The case of the longitudinal MF has been considered in details in \cite{MF-Arxiv,MF-SChFr} and it has been shown there that 	
soliton dynamics is a composition of a "free" soliton propagation along the molecular axis (i.e., as in the absence of the MF)  and oscillatory movement of an electron in the transverse direction, described by the functions of the harmonic oscillator 
with the frequency of oscillations and cyclotron mass
\begin{equation}
	\label{omega0}
	{{\omega }_{0}}=\frac{|{{B}_{0}}e\cos (\omega t)|}{m_{c}^{(l)}c}, \quad m_{c}^{(l)}=\sqrt{{{m}_{y}}{{m}_{z}}}.
\end{equation}

In the present paper we consider the case of a MF, perpendicular to the chain axis, since such a field has a qualitatively different impact on electron dynamics in the sense of its soliton features. Namely, we set 
\begin{equation}
	\label{B-tr}
	\vec{B}(t)=({0,{B}_{0}}\cos \omega t)
\end{equation}	 
and choose the following gauge invariance of the MF 
\begin{equation}
	\label{A-tr}
	\vec{A}=(0, 0, -{{B}_{0}}x\cos \omega t).
\end{equation}
In this case the Hamiltonian  can be written in the form $H= H_{s}+H_{tr }$, where 
\begin{equation}
	\label{Hperp}
	H_{s}=-\frac{{{\hbar }^{2}}}{2{{m}_{x}}}\frac{{{\partial }^{2}}}{\partial {{x}^{2}}}+\frac{1}{2{{m}_{z}}}{{\left( \hbar {{k}_{z}}+\frac{e}{c}{{B}_{0}}x\cos \omega t \right)}^{2}}
	-\sigma \rho (x,t),
\end{equation}
\begin{equation}
	\label{Htrl}
	H_{tr }=\frac{{{\hbar }^{2}}}{2{{m}_{y}}}\frac{{{\partial }^{2}}}{\partial {{y}^{2}}}
	+\frac{{{\hbar }^{2}}}{2{{m}_{z}}}\frac{{{\partial }^{2}}}{\partial {{z}^{2}}}.
\end{equation}
Hence, the wavefunction component  $ \psi _{tr}(y,z,t)$ is a normalized plane wave function in the $yz$-plane
\begin{equation}
	\label{psi-tr}
\psi_{tr }(y,z,t)=\frac{1}{\sqrt{L_yL_z}}e^{ik_yy+ik_zz-iE_{tr}t/\hbar}
\end{equation}
with $L_y,\;L_z$ being characteristic sizes of the system in $y$ and $z$ directions, respectively, while  the $x$-component of the wavefunction satisfies the nonlinear equation 
\begin{equation}
	\label{psi-perp}
	\left[ i\hbar \frac{\partial }{\partial t}+ \frac{\hbar^2}{2m_x}\frac{\partial ^2}{\partial x^2}+\sigma \rho(x,t)\right]\psi  (x,t)=\frac{1}{2m_x} \left[\hbar k_z + \frac{e}{c}B_0 x cos(\omega t)\right] \psi  (x,t),
\end{equation}
where the deformation of the chain is determined by the equation
\begin{equation}
	\label{eqrho2}
	\left( \frac{{{\partial }^{2}}}{\partial {{t}^{2}}}-{{V}_{0}}^{2}\frac{{{\partial }^{2}}}{\partial {{x}^{2}}} \right)\rho (x, t)+\frac{\sigma {{a}^{2}}}{M}\frac{{{\partial }^{2}}}{\partial {{x}^{2}}}{{\left| \psi (x, t) \right|}^{2}}=0.
\end{equation}

It is convenient to introduce the dimensionless units 
\begin{equation}
	\label{adim-par}
	\tilde{x}=\frac{\sqrt{g}}{a}x, \quad
	{{\tilde{x}}_{0}}=\frac{\sqrt{g}\hbar c{{k}_{z}}}{ae{{B}_{0}}}, \quad 
	\tau =2\frac{Jg}{\hbar }t,\quad \Omega =\frac{\hbar \omega }{2Jg}.
\end{equation}
and re-write  system of Eqs. (\ref{psi-perp}), (\ref{eqrho2}) as the modified NLS equation (see (\cite{MF-Arxiv,MF-SChFr})):
\begin{equation}
	\label{mNLS}
	\left[ i\frac{\partial }{\partial \tau }+\frac{1}{2}\frac{{{\partial }^{2}}}{\partial {{{\tilde{x}}}^{2}}}+\frac{1}{2}|{{\psi }}{{|}^{2}} \right]{{\psi }}
	=i\varepsilon R\left(\psi \right),
\end{equation}
where 
\begin{equation}
	\label{R}
	R\left(\psi \right)=-i\left[ {{\left( \tilde{x}\cos \Omega \tau +{{{\tilde{x}}}_{0}} \right)}^{2}}-\frac{\sigma }{2Jg}{{\rho }_{1}} \right]\psi ,
\end{equation}
\begin{equation}
	\label{eps}
	\varepsilon =\frac{{{e}^{2}}{{B}_{0}}^{2}{{a}^{2}}}{4{{m}_{z}}J{{c}^{2}}{{g}^{2}}}
\end{equation}
and $\psi \equiv \psi (\tilde{x},\tau ),\; \rho_1 \equiv \rho_1 (\tilde{x},\tau )$.

It can be easily checked that the coefficient (\ref{eps}) is small even for very strong available magnetic fields, $\varepsilon \ll 1$. Thanks to this, Eq. (\ref{mNLS}) can be solved using the perturbation method, based on the  inverse scattering technique for the NLS \cite{Karpman}, as it has been done in \cite{MF-SChFr}.\footnote {In spite of the fact of other approaches to solve  the nonlinear Schr\"odinger equation with time-periodic coefficients (see, e.g., references \cite{Bondila,Barashenkov,Friedland,Choi}), in the considered here problem the nonlinearity plays essential role, and, therefore,  we intend to avoid considering weakly nonlinear limit, but  use  the nonlinear perturbation theory \cite{Karpman}.}

Namely, it has been shown there that in the zero order approximation  the solution of Eq. (\ref{mNLS}) has a standard form
\begin{equation}
	\label{psi-sol0}
	\psi (\tilde{x},\tau )=\psi _{s}(\tilde{x},\tau )\equiv 2\nu \operatorname{sech}\zeta \exp \left[ i\varphi \right]
\end{equation}
in which
\begin{equation}
	\label{zeta}
	\zeta =2\nu \left( \tilde{x}-\tilde{\xi } \right), \qquad
	\varphi =\frac{\mu \zeta }{\nu }+\eta ,
\end{equation} 
with the constant amplitude, and, respectively, width: 
\begin{equation}
	\label{nu}
	\nu ={{C}_{1}}=\frac{\sqrt{g}}{4}, 
\end{equation}	
and other parameters slowly varying in time:
\begin{equation}
	\label{ksi}
	\tilde{\xi }=2\mu \tau +{{C}_{2}}, 
\end{equation}	
\begin{equation}
	\label{mu}
	\mu =-\varepsilon \biggl( \frac{{\tilde{\xi }}}{2\Omega }\left( \Omega \tau +\sin \Omega \tau \cos \Omega \tau  \right)
	-\frac{{{{\tilde{x}}}_{0}}}{\Omega }\sin \Omega \tau +\frac{2}{3}\nu \tau \alpha  \biggr)+{{C}_{3}}, 
\end{equation}	
\begin{equation}
	\label{eta}
	\eta =2\tau \left( {{\nu }^{2}}+{{\mu }^{2}} \right)+\varepsilon \left( \frac{{{\pi }^{2}}}{48\Omega {{\nu }^{2}}}-\frac{{{{\tilde{\xi }}}^{2}}}{\Omega } \right)\biggl( \frac{\Omega \tau }{2}
	+\frac{1}{2}\sin \Omega \tau \cos \Omega \tau  \biggr)+\frac{2\varepsilon \tilde{\xi }{{{\tilde{x}}}_{0}}}{\Omega }\sin \Omega \tau -\varepsilon \tau {{\tilde{x}}_{0}}^{2}+{{C}_{4}} .
\end{equation}
Here ${{C}_{1}}$, ${{C}_{2}}$, ${{C}_{3}}$, ${{C}_{4}}$ are constants of integration, and parameter $\alpha $ is introduced:
\begin{equation}
	\label{alpha}
	\alpha =\frac{2\nu {\sigma }^{2}dV/dt}{\varepsilon w J{{g}^{3/2}}({{V}_{0}}^{2}-{{V}^{2}})}.
\end{equation}

\section{Soliton radiation in MF}\label{deform}	

In the previous sections we have treated the dynamic equations in the adiabatic approximation and ignored the feedback of the MF induced electron parameters change  on the lattice deformation. In the present section, using the obtained above results, we will take this feedback into account and will show that this results in the change of the soliton dynamic mass, its dependence on the intensity and frequency of the MF and in soliton radiation at some resonant frequency. Thus, soliton  wave-function and lattice deformation in the self-consistent approach can be represented as the expansion
\begin{equation}	
	\label{psi-mf}
	\psi (x,t)={{\psi }_{s}}(u)+\varepsilon {{\psi }_{1}}(u), \qquad u=x-\xi (t)  
\end{equation}
\begin{equation}
	\label{rho-0-d}
	\rho(x,t)= 	\rho_0(x,t)-\frac{ \sigma }{w(1-s^2)}\left(\psi_s \psi^*_1+\psi_s ^*\psi_1\right)
	+\varepsilon \rho_1(x,t). 
\end{equation}
It can be easily checked that the zero order term of the lattice deformation  $\rho_0(u) $ has the same functional form as the deformation of the lattice in the absence of the MF (see Eq. (\ref{rho})), and is given by the expression 
\begin{equation}
	\label{rho-0-ad}
	\rho_0(x,t)=\frac{ \sigma }{w(1-s^2)}|\psi (x,t)|^2 \approx   \frac{ \sigma }{w(1-s^2)}|\psi _s(x,t)|^2 . 
\end{equation}
	
Taking into account that
\begin{equation}
	\label{deriv}
	\frac{\partial ^2}{\partial t^2}|\psi (x,t)|^2		
	=	\frac{\partial ^2}{\partial x^2}|\psi (x,t)|^2 \dot{\xi}^2-\frac{\partial }{\partial x}|\psi (x,t)|^2\ddot{\xi}, 
\end{equation}	
where  $\dot {\xi}$ means time derivative, in the first order with respect to the parameter $\varepsilon $ we get the equation
\begin{equation}
	\label{rho1-def}
	\frac{\partial ^2\rho _1(x,t)}{\partial t^2}-V^2_0\frac{\partial ^2\rho _1(x,t)}{\partial x^2}=
	-\frac{\sigma a}{w}\ddot{\xi}\frac{\partial }{\partial x}|\psi (x,t)|^2
\end{equation}		 
whose solution 	is given by the expression
\begin{equation}
	\label{rho1-def-s}
	\rho _1(x,t)=
	-\frac{\sigma a}{2\pi wV_0}\int_{-\infty}^{t}dt'\ddot{\xi}(t')
	\int_{-\infty}^{\infty}dq\frac{\sin\left[V_0q\left(t-t'\right)\right]}{q}
	\int_{-\infty}^{\infty}dx'e^{iq(x-x')}
	\frac{\partial }{\partial x'}|\psi (x',t')|^2.
\end{equation}	

It is easy to see that time derivatives of the soliton center of mass coordinate (see Eq. ({\ref{psi-sol0}})) are 
\begin{equation}
	\label{cmc}
	\dot{\xi}=\frac{\hbar}{m_x}\mu ,\qquad 	\ddot{\xi}=\frac{\hbar}{m_x}\dot{\mu },
\end{equation}
where, according to Section \ref{MF},
\begin{equation}
	\label{cmc1}
	\dot{\mu}=-\varepsilon \left(\tilde{\xi }\cos ^2 \omega t+\tilde{x }_0\cos  \omega t
	+\frac{2wJg}{\sigma}\int_{-\infty}^{\infty}\rho_1 (\zeta)	\frac{\sinh \zeta}{\cosh ^3 \zeta} d\zeta 
	\right).
\end{equation}	
These results show that the equation for the soliton center of mass coordinate has the form of the Newton equation for a particle in the presence of the periodic force $F(t)=F_0\sin (\omega t)$:
\begin{equation}
	\label{newt}
	m_x\ddot{\xi}(t)+\int_{0}^{t}\ddot{\xi}(t-t')Q(t')dt'=F_0\sin \omega t
\end{equation}
in which
\begin{equation}
	\label{Q}
	Q(t)=\frac{\sigma^2 a \nu}{V^2_0}\frac{d}{dt}\left[\frac{\nu V_o t\cosh\left(\nu V_0t\right) -\sinh \left(\nu V_0t\right)}{\sinh ^3\left(\nu V_0t\right)}
	\right]
\end{equation}
\begin{equation}
	\label{F0}
	F_0=e\omega B_0 x.
\end{equation}
Equation (\ref{newt}) can be written in the form
\begin{equation}
	\label{xitt}
	\ddot{\xi}(t)=\frac{F_0}{m_{dyn}(\omega)}\sin\left(\omega t-\phi(\omega)
	\right)
\end{equation}
where $m_{dyn}(\omega) $,  soliton dynamic mass, and $ \phi(\omega)$,  phase shift of soliton oscillations as comparing with the external force, are  determined by the kernel of the integral  in Eq. (\ref{newt}), and, thus, are functions of the frequency of the MF. Calculation of the explicit expressions of $m_{dyn}(\omega) $ and  $ \phi(\omega)$ are very similar to calculations of these parameters for the case of soliton dynamics in external electromagnetic field in \cite{BC-HE} and we omit them here, reminding that it has been shown there that the dynamic mass of a soliton is a non-monotonous function of $\omega$  with two different functional dependencies at $ \omega < \omega_{res}$ and $ \omega > \omega_{res}$, where $\omega_{res} $ is determined by the expression
\begin{equation}
	\label{omega-res}
	\omega_{res} =\frac{gV_0}{\pi a}.
\end{equation}
It follows from Eq. (\ref{omega-res}) that $\omega_{res} $ is characteristic soliton frequency determined by the time needed for the sound waves to pass the soliton width, $l_s=\pi a/g$. 
At $ \omega \ll \omega_{res}$ soliton dynamic mass coincides with the effective mass of a free soliton, while in fast oscillating MFs, at $ \omega \gg \omega_{res}$, soliton dynamic mass coincides with the effective mass of an electron, since the lattice deformation due to the shift of heavy atoms from their equilibrium can't follow fast oscillations of light electron, although electron still remains in a bound soliton state. 

Another important consequence of the self-consistent account of the lattice deformation is soliton radiation due to the term $\rho_1(x,t)$ which is given by the expression (\ref{rho1-def-s}). Calculating the first derivative of the soliton wavefunction $\psi(x,t)\approx \psi_s (x,t)$ where $\psi_s (x,t)$ is defined in Eq. (\ref{psi-sol0}), we can  perform the integration in (\ref{rho1-def-s}):
\begin{equation}
	\label{int1}
	\int_{-\infty}^{\infty}	dx'e^{iq(x-x')}
	\frac{\partial }{\partial x'}|\psi _s(x',t')|^2=-8\nu^2 e^{-iq(x-\xi)}\left(1-\frac{i\pi q^2}{8\nu^2}\sinh ^{-1}\frac{q\pi}{4\nu}\right),
\end{equation}
so that
\begin{equation}
	\label{int2}
	-8\nu^2 \int_{-\infty}^{\infty}dq\frac{\sin\left[V_0q\left(t-t'\right)\right]}{q}
	e^{-iq(x-\xi)}\left(1-\frac{i\pi q^2}{8\nu^2}\sinh ^{-1}\frac{q\pi}{4\nu}\right)=-\pi \left[
	|\psi_s(\xi_+)|^2-|\psi_s(\xi_-)|^2
	\right]
\end{equation}
where the notation is used
\begin{equation}
	\label{asymp}
	|\psi_s(\xi_{\pm})|^2=4\nu^2 \cosh ^{-2}\left[2\nu (\tilde{x}-\tilde{\xi }\pm V_0 \tau).
	\right]
\end{equation}
Substituting this result in Eq. (\ref{rho1-def-s}), we get
\begin{equation}
	\label{rho1-2}
	\rho_1(x,t)=-\frac{\sigma a}{2 wV_0}\int_{0}^{t}d\tilde{\tau}\ddot{\tilde {\xi}}(t-\tilde {\tau})\left[
	|\psi_s(\xi_+)|^2-|\psi_s(\xi_-)|^2
	\right],
\end{equation}
and, taking into account relation (\ref{asymp}),
\begin{equation}
	\label{rho1-3}
	\rho_1(x,t)=-\frac{\sigma a}{2 wV_0}e\omega B_0\int_{0}^{t}d\tau [\sin \left[\omega\left(t-\tau
	\right)-\phi(\omega)
	\right] \left[
	|\psi_s(\xi_+)|^2-|\psi_s(\xi_-)|^2
	\right].
\end{equation}
This integral can't be taken analytically, and, instead of calculating it numerically for arbitrary $x$ and $t$, we can calculate its asymptotics far from the soliton center of mass. Thus, we obtain the asymptotic expressions for the radiated sound waves in the form of oscillations with the frequency of the MF oscillations:
\begin{equation}
	\label{rho+}
	\rho_1(x,t)=A(\omega)\cos \left[\omega \left(t-\frac{x-\xi}{V_0}-\phi \right)\right], \quad x-\xi \rightarrow \infty ,
\end{equation}
\begin{equation}
	\label{rho-}
	\rho_1(x,t)=-A(\omega)\cos \left[\omega \left(t+\frac{x-\xi}{V_0}-\phi \right)\right], \quad x-\xi \rightarrow -\infty .
\end{equation}
The amplitude of the radiated waves $A(\omega)$ depends on the soliton parameters and frequency of the MF:
\begin{equation}
	\label{Ampl}
	A(\omega)=\frac{a\sigma \omega F_0}{2m_{dyn}wV^2_0\omega_{res}}\sinh ^{-1}\left(\frac{\omega}{\omega_{res}}\right)
\end{equation}
where $\omega_{res}$ is determined in Eq. (\ref{omega-res}). It is worth to stress that this characteristic frequency which determines the resonant regime of the perpendicular MF impact on soliton dynamics is different from the cyclotron soliton frequency in the parallel MF given in Eq. (\ref{omega0}).

\section{Account of energy dissipation}\label{dissip}

According to the obtained above results, soliton velocity in an oscillating MF is a function of time: $V(\tau) = d\xi / d\tau $ (see Eqs. (\ref{ksi}), (\ref{mu})) and the soliton moves with an acceleration, $ dV(\tau)/ d\tau \neq 0$, which is possible due to absorption of the energy from the external field. In the adiabatic approximation in solution (\ref{psi-sol0}) we have neglected changes of the soliton envelope form and increasing in time its oscillating tails (linear waves). First of all, we know that in the harmonic approximation of the lattice interactions, the soliton velocity is limited by the velocity of the sound in the chain, therefore, at large enough velocities we have to take into account the anharmonicity of the lattice vibrations; at even bigger  velocities soliton radiation of sound waves increases and adiabatic approximation becomes not valid. 

Nevertheless, in real systems there is always present dissipation of energy. Very often energy dissipation plays negative role. But the situation can be different when we consider nonlinear systems, such as our case. Due to the energy dissipation we can expect deceleration of soliton in MF. Indeed, in real situations molecular chains are surrounded by some medium, the chains themselves have a complex structure and in their phonon spectrum there are present several phonon modes, so that one has to take into account energy exchange between electron and phonon subsystems with the medium and other degrees of freedom. For qualitative study it is convenient to introduce the friction by adding the corresponding term in Eq. (\ref{mNLS}) (see \cite{LB-fr})
\begin{equation}
	\label{mNLS-fric}
	\left[ i\frac{\partial }{\partial \tau }+\frac{1}{2}\frac{{{\partial }^{2}}}{\partial {{{\tilde{x}}}^{2}}}+\frac{1}{2}|{{\psi }}{{|}^{2}} \right]{{\psi }}
	=i\varepsilon R\left(\psi\right) +V_{fr}\left(\psi\right)\psi
\end{equation}
where 
\begin{equation}
	\label{V-fr}
V_{fr}\left(\psi\right)= -i\gamma \frac{\hbar}{2}\ln \frac{\psi}{\psi^*},
\end{equation}
 $\gamma$ is the friction coefficient and the sign $^*$ means the complex conjugation. We also set that $ \psi(\tilde{x},\tau )=\psi_{s}(\tilde{x},\tau )$ at $\tau =0$, where $ \psi_{s}(\tilde{x},\tau )$ is determined in Eq. (\ref{psi-sol0}). Therefore, Eq. (\ref{mNLS-fric}) can be rewritten in the form 
 \begin{equation}
 	\label{mNLS-fr}
 	\left[ i\frac{\partial }{\partial \tau }+\frac{1}{2}\frac{{{\partial }^{2}}}{\partial {{{\tilde{x}}}^{2}}}+\frac{1}{2}|{{\Psi }}{{|}^{2}} \right]{{\Psi }}
 	=i\varepsilon R_{fr}\left(\psi\right) 
 \end{equation}
 where 
\begin{equation}
\label{R-fr}	
 R_{fr}\left(\psi \right)= 	-i\left[ {{\left( \tilde{x}\cos \Omega \tau +{{{\tilde{x}}}_{0}} \right)}^{2}}-\frac{\sigma }{2Jg}{{\rho }_{1}} \right]\psi -\frac{\gamma}{4Jg\varepsilon}\psi \ln \frac{\psi}{\psi^*}.
\end{equation} 
 
This equation similar to Eq. (\ref{mNLS}) can be solved using the perturbation method \cite{Karpman}. To avoid extra complications, in this section we will ignore in the lattice deformation the feedback of the change of the solitob wavefunction envelope which leads to the radiation of waves considered in the previous section, and set 
\begin{equation}
	\label{rho-0-nad}
	\rho(x,t)= 	\rho_0(x,t)	+\varepsilon \rho_1(x,t) 
\end{equation}
where $\rho_1$ satisfies the equation
\begin{equation}
	\label{rho-1}
	\frac{d{{\rho }_{1}}}{du}=-\frac{q}{\varepsilon a {{V}_{0}}^{2}\left( 1-s^{2} \right)}{{\rho }_{0}}, \qquad 	q=a\frac{dV}{dt}.
\end{equation}
In this adiabatic approximation 
we have the soliton wavefunction in the form (\ref{psi-sol0}), (\ref{zeta}) with the parameters $\nu,\;\mu,\; \tilde{\xi},\; \eta$ depending on time. Substituting the explicit expression (\ref{R-fr}) for the term in the right hand side of Eq. (\ref{mNLS-fr}) into the equation which determine time dependence of soliton parameters accoording to the perturbation method, we derive the following equations
\begin{equation}
	\label{nu-fr}	
\frac{d\nu	}{d\tau}=0, 
\end{equation} 
\begin{equation}
	\label{ksi-fr}	
	\frac{d	\tilde{\xi}}{d\tau}=2\mu, 
\end{equation} 
\begin{equation}
	\label{eta-fr}	
	\frac{d\eta	}{d\tau}= -\varepsilon \left[ \tilde{\xi}\cos ^2 \omega\tau +\tilde{x}_0 \cos  \omega\tau +\Gamma \mu \right]
\end{equation} 
\begin{equation}
	\label{mu-fr}	
	\frac{d\mu	}{d\tau}=2(\nu^2+\mu^2)+\varepsilon \left[ \left(\frac{\pi^2}{48\nu^2} - \tilde{\xi}^2 \right)\cos^2 \omega\tau - 2\tilde{\xi} \tilde{x}_0  \cos  \omega\tau - \tilde{x}_0 ^2-\Gamma  \eta	
	\right]
\end{equation} 
where 
\begin{equation}
	\label{Gamma}	
\Gamma=	\frac{\gamma }{2Jg\varepsilon}.
\end{equation} 

It is easy to see that the soliton has constant amplitude, and, respectively, width, as in the absence of the friction: 
\begin{equation}
	\label{nu-sol-fr}
	\nu ={{C}_{1}}=\frac{\sqrt{g}}{4}, 
\end{equation}	
while its other parameters are slowly varying in time and contain terms oscillating in time with the frequency of the MF:
\begin{equation}
	\label{ksi-rad}
	\tilde{\xi }(\tau)=2\mu \tau +{{C}_{2}}, 
\end{equation}	
\begin{equation}
	\label{mu-rad}
	\mu (\tau) =-\varepsilon \biggl( \frac{{\tilde{\xi }}}{2\Omega }\left( \Omega \tau +\sin \Omega \tau \cos \Omega \tau  \right)-
	\frac{{{{\tilde{x}}}_{0}}}{\Omega }\sin \Omega \tau +A \tau   \biggr)+{{C}_{3}}, 
\end{equation}
\[
\eta (\tau) =2\tau \left( {{\nu }^{2}}+{{\mu }^{2}} \right)+\varepsilon \left( \frac{{{\pi }^{2}}}{48\Omega {{\nu }^{2}}}-\frac{{{{\tilde{\xi }}}^{2}}}{\Omega } \right)\biggl( \frac{\Omega \tau }{2}
+\frac{1}{2}\sin \Omega \tau \cos \Omega \tau  \biggr)+
\]	
\begin{equation}
	\label{eta-rad}
	\qquad + \frac{2\varepsilon \tilde{\xi }{{{\tilde{x}}}_{0}}}{\Omega }\sin \Omega \tau -\varepsilon \tau {{\tilde{x}}_{0}}^{2}+{{C}_{4}} .
\end{equation}
Here parameter $A $ is introduced:
\begin{equation}
	\label{A}
	A = \frac{8}{3} \nu ^2 \frac{q}{\varepsilon \sqrt{g}{V}_{0}^{2}}
\end{equation}
and ${{C}_{1}}$, ${{C}_{2}}$, ${{C}_{3}}$, ${{C}_{4}}$ are constants of integration. From the normalization condition of the wave function we get $C_1=\sqrt{g}/4$ and other constants  can be set equal to zero.

	\section{Conclusions}

Thus, we have shown that in the external oscillating in time MF perpendicular to molecular chains orientation, soliton acquires  complex dynamics. In particular, its velocity, acceleration, generalized momentum and phase are slowly varying in time functions which contain oscillating in time terms  with the frequency of the MF and its higher harmonics.  Due to acceleration, soliton velocity does not increase infinitely, its increase is suppressed by the energy dissipation, since, according to Eqs. (\ref{ksi-fr}) and (\ref{mu-fr}), soliton velocity riches its asymptotic value 
\begin{equation}
	\label{vel-as}
	\dot{\xi}=2\mu 
\end{equation}
and
\begin{equation}
	\label{accel-as}
	\ddot{\xi}=-2\varepsilon \left(\tilde{\xi}^2\cos^2\Omega \tau+ \tilde{x}_0\cos\Omega \tau + \Gamma \mu_0 e^{-\varepsilon \Gamma \tau} 
	\right). 
\end{equation}

The soliton center of mass which has the meaning of the collective spacial variable, is oscillating in time function with the frequency of oscillations equal to the MF frequency and its higher harmonics. This processes is accompanied by the radiation of the sound waves of small amplitude $\varepsilon \rho_1 (x,t)$, $\varepsilon \ll 1$ in both directions from the soliton center of mass  given by expressions (\ref{rho+}), (\ref{rho-}) whose amplitude depends on parameters of the molecular chain and  field parameters. This process is the most intensive at the resonant frequency of the MF, $\omega = \omega_{res}$, where $\omega_{res}$ is characteristic soliton frequency (\ref{omega-res}), different from the soliton cyclotron frequency in the parallel MF. Obtained results show that soliton dynamics has two different regimes at low and high frequencies of the magnetic field as comparing with the characteristic soliton frequency.

Thus, we conclude that such complex impact of the oscillating in time MF on soliton dynamics essentially affects the conductivity of low-dimensional systems which support existence of solitons. 
Thus, our results show that oscillating MFs can significantly affect charge transport processes in micro- and nanoelectronic devices, in the devices of biomimetic technologies that are based on polymers and other quasi-one-dimensional materials (see, e.g., \cite{Sirri,Yao,Shelte,Ahmad} and references therein), in the therapeutic and diagnostic devices based on the so called "next generation" low-dimensional materials (see \cite{Minamiki,Chen2,Zhou}).

There is also one more important consequence of the obtained results. It concerns biological systems: we have shown that external alternating MFs, changing the dynamics of solitons, modify  charge and energy transport in the redox processes in biological systems. Therefore, we can expect that the obtained results can explain the physical mechanism of the resonant therapeutic effects of low intensity oscillating MFs \cite{BrizhikFermi,LiboffSmith,Anderson,Emre,Funk,EMEF}, since the most distinct modification of soliton dynamics occurs at the resonant frequency of the MF, $\omega = \omega_{res}$. It is worth to stress here, that oscillating in time character of soliton dynamics is accompanied first of all by the excitation of the vibrations  of polypeptide chain side groups and  radiation of the sound waves, which as it is well known, is very important from the point of view of structure--function relations and recognition in biological systems. Excitation of bending vibrations is by itself important for the formation of localized excitations \cite{Kovalev-double}. Secondly, such character of the charged soliton dynamics, according to the Maxwell equations, is accompanied by the radiation of the electromagnetic waves (see, e.g., \cite{BrHoMu}) which can play the regulating role in the physiological processes in living systems, as is also confirmed by some clinical studies, e.g., \cite{Adamski}.

Finally, we underline, that using of the perturbation method to solve the dynamic equations for soliton in the MF presence does not restrict the validity of the obtained results for large time scales (see also \cite{Kovalev-asymp}) and for strong MFs in view of real smallness of the parameter $\varepsilon $. For instance, for the geomagnetic field whose intensity is $B_0=30-50 \; \mu $Tl, for characteristic values of the polypeptide parameters \cite{Scott}, the small parameter (\ref{eps}) is $\varepsilon =10^{-34}$, and even for magnetic fields used in tomography, $B_0=1$ Tl, $\varepsilon =1.26 \cdot 10^{-24}$. Nevertheless, as we have shown, even very weak MFs essentially modify soliton dynamics and, therefore, charge transport processes in the low-dimensional systems.

	\vskip5mm 
	{\bf Acknowledgement.} 
	\textit{The author expresses thanks to K. Temchenko for fruitful discussions during her work on the master degree thesis. This work was supported by the fundamental scientific program 0122U000887 of the Department of Physics and Astronomy of the National Academy of Sciences of Ukraine. The author acknowledges also the Simons Foundation.}

\end{document}